\documentclass[pdflatex,sn-mathphys]{sn-jnl}

\jyear{2022}%

\theoremstyle{thmstyleone}%
\theoremstyle{thmstyletwo}%

\theoremstyle{thmstylethree}%

\begin{document}

\title[ ]{Direct Observation of a Superconducting Vortex Diode}


\author*[1]{\fnm{Alon} \sur{Gutfreund}}\email{alon.gutfreund@mail.huji.ac.il}

\author[2]{\fnm{Hisakazu} \sur{Matsuki}}

\author[3]{\fnm{Vadim} \sur{Plastovets}}
\author[1]{\fnm{Avia} \sur{Noah}}

\author[2]{\fnm{Laura} \sur{Gorzawski}}
\author[1]{\fnm{Nofar} \sur{Fridman}}
\author[2]{\fnm{Guang} \sur{Yang}}
\author[3]{\fnm{Alexander} \sur{Buzdin}}
\author[1]{\fnm{Oded} \sur{Millo}}
\author*[2]{\fnm{Jason} W. A.  \sur{Robinson}\email{jjr33@cam.ac.uk}}
\author*[1]{\fnm{Yonathan} \sur{Anahory}\email{yonathan.anahory@mail.huji.ac.il}}

\affil*[1]{\orgdiv{The Racah Institue of Physics}, \orgname{The Hebrew University of Jerusalem},\orgaddress{ \postcode{9190401}, \country{Israel}}}

\affil[2]{\orgdiv{Department of Materials Science \& Metallurgy}, \orgname{University of Cambridge}, \orgaddress{\city{Cambridge}, \postcode{CB3 0FS}, \country{United Kingdom}}}

\affil[3]{\orgname{University of Bordeaux}, \orgaddress{\city{Talence}, \postcode{LOMA UMR-CNRS 5798, F-33405}, \country{France}}}

\abstract{The interplay between magnetism and superconductivity can lead to unconventional proximity and Josephson effects. A related phenomenon that has recently attracted considerable attention is the superconducting diode effect, in which a non-reciprocal critical current emerges.
Although superconducting diodes based on superconducting/ferromagnetic (S/F) bilayers were demonstrated more than a decade ago, the precise underlying mechanism remains unclear. While not formally linked to this effect, the Fulde-Ferrell-Larkin-Ovchinikov (FFLO) state is a plausible mechanism, due to the 2-fold rotational symmetry breaking caused by the finite center-of-mass-momentum of the Cooper pairs. Here, we directly observe, for the first time, a tunable superconducting vortex diode in Nb/EuS (S/F) bilayers. Based on our nanoscale SQUID-on-tip (SOT) microscope and supported by \textit{in-situ} transport measurements, we propose a theoretical model that captures our key results. Thus, we determine the origin for the vortex diode effect, which builds a foundation for new device concepts.}


\maketitle

\section*{Introduction}\label{introduction}
Conventional (s-wave) superconductivity is mediated by spin-singlet Cooper pairs in which each electron of a pair has the opposite sign of spin. Ferromagnetism favours a parallel alignment of electron spins and so the proximity effect at a superconductor/ferromagnet (S/F) interface suppresses superconductivity \citep{Bergeret2001,buzdin_review,Eschrig2008,Robinson2010,Anwar2010,Khaire2010,SC_spintronics,Yao2021,Komori2020:ScienceAdvances,Boost,PhysRevX.10.031020,PhysRevB.99.024507,PhysRevB.99.144503} and breaks time reversal symmetry, establishing the Fulde-Ferrell-Larkin-Ovchinikov (FFLO) state \cite{fulde_ferrell,larkin_ovchinnikov}.

In a system that exhibits Rashba spin-orbit-coupling (SOC)\citep{Rashba_SOC}, the spin bands split, and a ferromagnetic exchange field acts differentially on the oppositely-aligned electron spins within the Cooper pairs \cite{moodera_exchange_field}. Hence the Cooper pairs gain a non-zero centre-of-mass-momentum that is magnetization-orientation-dependent \citep{moodera_symmetry_breaking,Rashba_SC_diode_theory}. This type of symmetry breaking has been observed in S/F bilayers as a non-reciprocal critical current \citep{asymmetric_Ic_barber,Jeon2022,moodera_diode_effect,Suri2022,josephson_diode_effect,SC_diode_effect,Baumgartner2022,Wu2022} and an asymmetric intermediate-state of resistance \citep{CGC_PyNb,CSC_vortex_dynamics}. Despite recent theoretical progress\citep{Rashba_SC_diode_theory}, the connection of these observations to the microscopic explanation remains unclear.
\par
In the present study, we focus on the origin of the diode effect in bilayers of EuS/Nb. Using magnetic imaging techniques, we demonstrate asymmetric vortex dynamics that are manipulated by the magnetization of the ferromagnet and correspond with a non-reciprocal critical current. While the role of vortex dynamics is largely overlooked in the literature regarding the diode effect, the unidirectional trajectories of vortex motion that we observe suggest an alternative underlying mechanism. These surprising results can be explained by taking into account the screening current distribution, induced by an inhomogeneous magnetic field emanating from the ferromagnet.
\newline
\section*{Results}
\begin{figure}[h]
    \centering
    \includegraphics[width=1\textwidth]{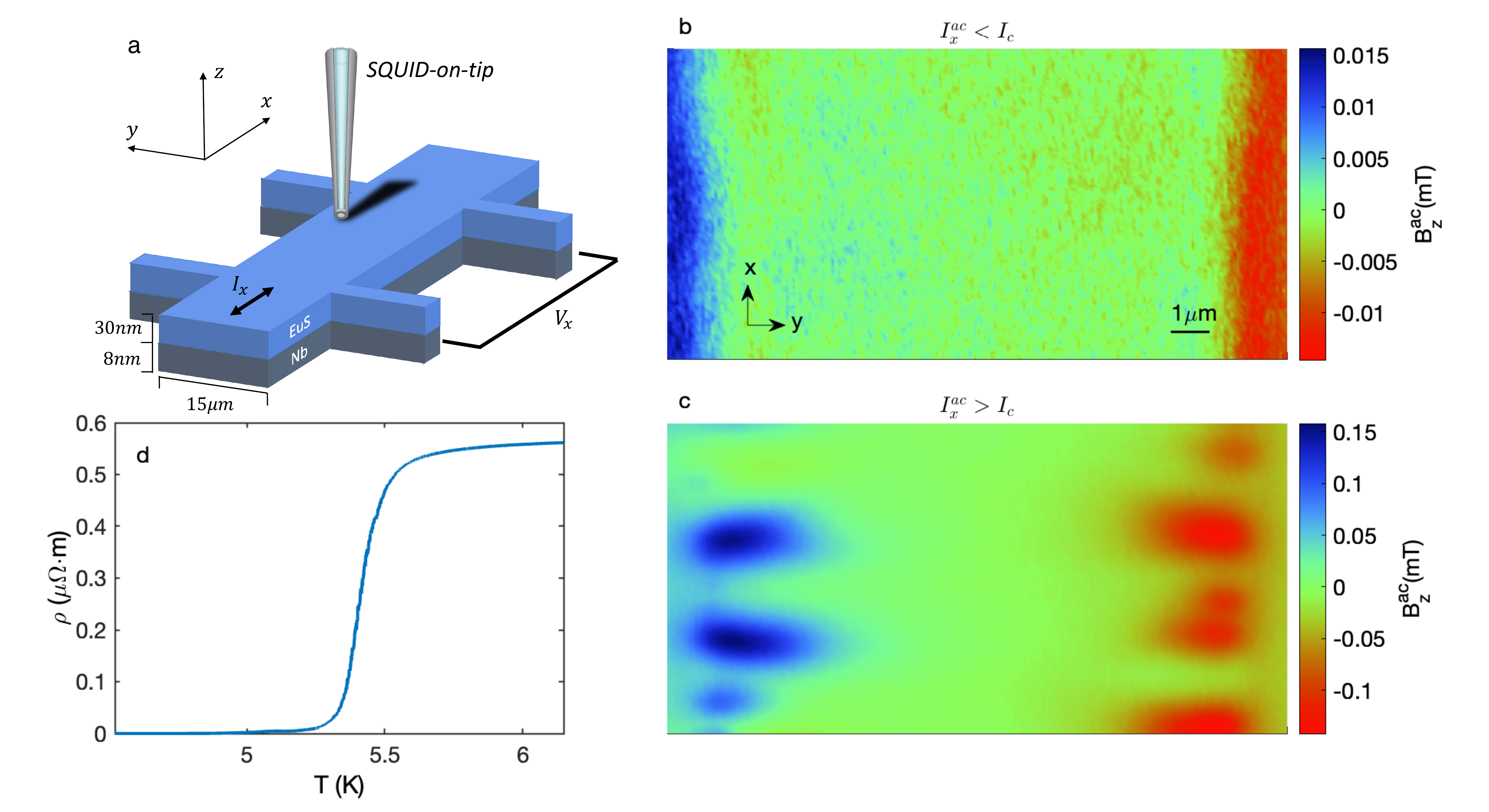}
    \caption{\small \textbf{Magnetic response of the EuS/Nb bilayer with an ac current $I_x^{ac}$}. (\textbf{a}) Schematic diagram of the measurement setup, showing a EuS/Nb Hall bar structure, along with the scanning SQUID-on-tip (SOT) probe. (\textbf{b}) SOT image of the ac out-of-plane component of the magnetic field $B_z^{ac}(x,y)$ modulated with an oscillating transport current amplitude $I_x^{ac}<I_c$. Blue(red) corresponds to a positive(negative) out-of-plane component of the field emanating from the Nb strip. The device was zero-field cooled to 4.2 K, below the superconducting transition ($T_c\sim 5.5 K$) and Curie temperature ($T_C\sim 20 K$). (\textbf{c}) same as \textbf{b} but with $I_x^{ac}>I_c$. In this case the polarity of the signal depends on whether the magnetic feature appears in phase (blue) or at a $\pi$-phase (red) with respect to the oscillating current. (\textbf{d}) $R(T)$ measurements of the device showing the superconducting transition}
    \label{fig:System&Current_dependance}
\end{figure}

The device consists of a EuS/Nb bilayer patterned into a Hall bar (see Fig. \ref{fig:System&Current_dependance}a and methods). We assume, that there should be no effect on the magnetic properties of the EuS film due to the coupling with superconducting Nb. We verified this assumption by performing global magnetization measurements on an unpatterned EuS/Nb device with identical thicknesses, which show a negligible difference across the SC transition temperature (supplementary Fig. \ref{fig:MH_MT}). We apply an ac current $I^{ac}_x$ along the EuS/Nb wires and simultaneously record the magnetic field response that is modulated in phase with the current using the SOT. In order to generate free vortices in the sample an OOP field $\lvert \mu_0H_z\rvert=5$ mT was applied in all the measurements presented in this work.
In Fig. \ref{fig:System&Current_dependance}b we show a $B^{ac}_z(x,y)$ image that corresponds to a longitudinal voltage $V_x=0$, meaning that the current amplitude is smaller than the critical current $I_c$. The image shows the expected features for a Biot-Savart field induced by a current flowing in a superconducting slab \citep{Zeldov_current_distribution,zeldov_flux_flow}. For $I^{ac}_x>I_c$  ($V_x\neq0$), lobe-shaped features appear in the image at the sample edges (see Fig. \ref{fig:System&Current_dependance}c). We note that a positive signal (blue on the colormap), means that the magnetic feature, in the time domain, appears in phase with respect to the applied current, whereas a negative signal (red on the colormap) means that there is a $\pi$ phase difference with respect to the current. Since these features appear with the onset of voltage, it is reasonable to assume that they are the result of vortex flow. Such vortex channels were already observed by applying a dc current \citep{zeldov_flux_flow}. The channels that are in-phase ($\pi$-phase) with respect to the current appear when the instantaneous current is $I_x^{ac}(t)>0$ ($I_x^{ac}(t)<0$).
\newline
Once the vortices penetrate deeper into the sample, they bifurcate and can take many distinct routes \citep{zeldov_flux_flow}. Given that we average over an ac signal with a time scale many orders of magnitude larger than those involved in the vortex dynamics, our images effectively portray the density of vortex paths across the device. For that reason, the signal along the channels becomes undetectable in the inner part of the sample. We emphasize that we are observing vortices in a thin film superconductor with an effective penetration depth determined by the Pearl length, $\Lambda\sim 6   $ $\mu m$. This length scale defines the effective size of the vortex. Having a larger vortex implies that the same magnetic flux is spread out over a larger area, resulting in weaker fields that are harder to detect. This further smears the vortex signal, hampering the observation of single vortex channels.
\begin{figure}[h]
    \centering
    \includegraphics[width=1\textwidth]{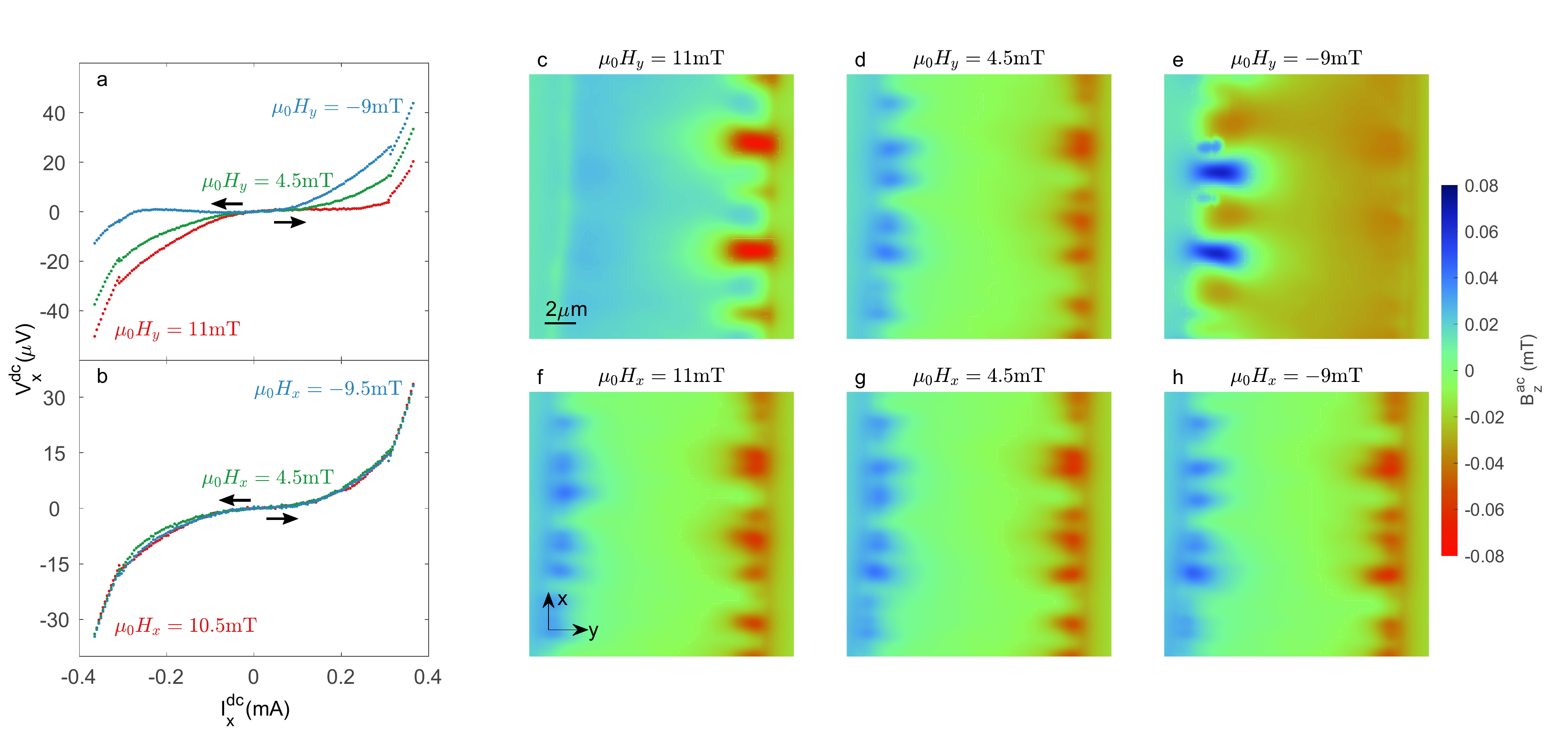}
    \caption{\small \textbf{Vortex flow and corresponding transport measurements as a function of an applied in-plane magnetic field.} (\textbf{a}-\textbf{b}) $I(V)$ characteristics for different transverse \textbf{a} and longitudinal \textbf{b} magnetic fields, arrows indicate the direction of current sweep. The red and blue curve are at fields beyond the saturation field ($H_s$) while the green curve is at the coercive field. Black arrows indicate that the transport curves were always swept from zero to maximum bias, in order to eliminate the effect of hysteresis caused by heating.(\textbf{c}-\textbf{h}) SQUID-on-tip (SOT) image of the ac out-of-plane component of the magnetic field $B_z^{ac}(x,y)$ modulated  with respect to an oscillating transport current $I_x^{ac}>I_c$. The polarity of the signal depends on whether the magnetic feature appears in phase (blue) or at a $\pi$-phase (red) with respect to the oscillating current. (\textbf{c}-\textbf{e}) Transverse magnetic field orientation ($H_y$) with $\mu_0H_y$ above $H_s$ in the $+y$ direction \textbf{c}, $-y$ direction \textbf{e} and at the coercive field \textbf{d}. (\textbf{f}-\textbf{h}) Same values of the magnetic field but in a longitudinal orientation, parallel to the direction of current $H_x \parallel I_x^{ac}$. For a full set of $B_z^{ac}(x,y)$ images with a transverse field orientation ($H_y$), see \href{https://youtu.be/i7FJTW6-7vQ}{supplementary movie 1}. For  a longitudinal field orientation ($H_x$), see \href{https://youtu.be/__iWGTqikUA}{supplementary movie 2}}
    \label{fig:AC_images_field_dependant}
\end{figure}

We first discuss the case where we fully magnetize the EuS by applying an in-plane magnetic field in the $y$ direction ($\Vec{H}\parallel\hat{y}$), perpendicular to the current flow, $H_y=11$ mT. The magnetic image $B^{ac}_z(x,y)$ acquired in that state is shown in Fig. \ref{fig:AC_images_field_dependant}c. Unlike what is observed in the \textit{Zero-Field-Cooled} (ZFC) state (\ref{fig:System&Current_dependance}c), the vortex channels are now visible only on the right edge of the sample. Interestingly, once the EuS is fully magnetized in the $-y$ direction, by applying $H_y=-9$ mT, vortices penetrate only from the opposite, left edge (Fig. \ref{fig:AC_images_field_dependant}e). This suggests that as a result of the sample magnetization, vortices only penetrate into the sample from one edge. Around the coercive field ($H_c=4.5$ mT), the anti-symmetric image observed in the ZFC state is recovered (Fig. \ref{fig:AC_images_field_dependant}d). These results clearly demonstrate the emergence of a vortex diode effect in which the diode direction is set by the magnetization and vanishes around $H_c$.

We now turn to discuss the case where the magnetization is in the $x$ direction ($\Vec{H}\parallel\hat{x}$), parallel to the current. In this case, the vortex channels appear on both edges regardless of whether the magnetization is in the $+x$ direction (Fig. \ref{fig:AC_images_field_dependant}f), $-x$ direction (Fig. \ref{fig:AC_images_field_dependant}h), or at the coercive field (Fig. \ref{fig:AC_images_field_dependant}g). These results indicate that the vortex diode effect vanishes when the magnetization is parallel to the current or that there is no net magnetization.
\newline
If the vortices enter with more ease from one edge compared to the other, it implies that the critical current in the positive direction should be different from that in the negative direction ($\lvert I^+_c\rvert\neq \lvert I^-_c\rvert)$. To confirm the absence of a reciprocal critical current, we pass a dc current $I^{dc}_x$ while measuring the longitudinal voltage $V_x$ for different magnetization directions. The results for the case where the in-plane magnetization is perpendicular to the current are depicted in Fig. \ref{fig:AC_images_field_dependant}a. The red curve was measured while applying $H_y=11$ mT and corresponds to the SOT image shown in Fig. \ref{fig:AC_images_field_dependant}c. A clear asymmetry is observed between the sweep from zero to maximum positive current compared to the opposite direction, in accordance to the unidirectional vortex flow image shown in Fig. \ref{fig:AC_images_field_dependant}c. The asymmetry is reverted by changing the magnetization direction (blue curve, acquired under $H_y=-9$ mT), consistent with the reversed vortex flow direction (Fig. \ref{fig:AC_images_field_dependant}e). Importantly, no asymmetry in the $I(V)$ characteristics is observed around the coercive field (green curve), consistent with vortices penetrating from both edges (Fig. \ref{fig:AC_images_field_dependant}d). In the case where the field is applied along the $x$ axis (\ref{fig:AC_images_field_dependant}b), the curves are nearly independent of the applied field and give nearly symmetrical values of $I_c$, consistent with the images (Fig. \ref{fig:AC_images_field_dependant} f-h).
\begin{figure}[h]
    \centering
    \includegraphics[width=1\textwidth]{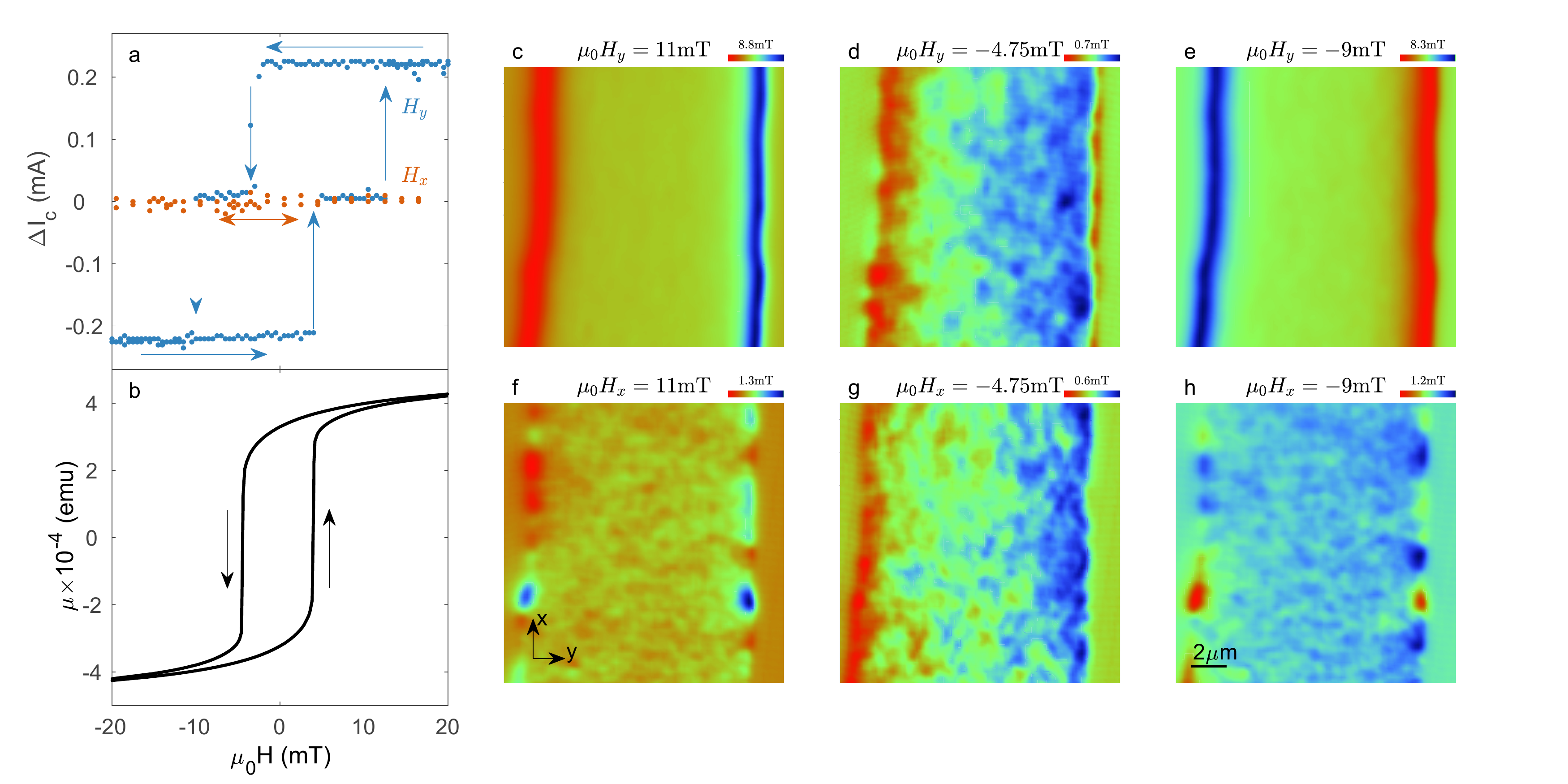}
    \caption{\small \textbf{Correlation between the vortex diode effect and the magnetic texture of EuS.} (\textbf{a}) asymmetry factor $\Delta I_c=\lvert I_c^+\rvert-\lvert I_c^-\rvert$ as a function of magnetic field for transverse (blue symbols) and longitudinal (red symbols) magnetization. The arrows mark the magnetic field sweep directions. (\textbf{b}) in-plane M(H) curve of an unpatterned EuS/Nb film with the same thicknesses as our device. (\textbf{c-h}) SQUID-on-tip (SOT) images of the static out-of-plane component of the magnetic field $B_z^{dc}(x,y)$ emanating from the EuS/Nb bilayer under various applied in-plane magnetic fields, as indicated. (\textbf{c-e}) Transverse ($H_y$) magnetic field orientation with $\mu_0H_y$ above the saturation field in the $+y$ direction \textbf{c}, $-y$ direction \textbf{e} and at the coercive field \textbf{d}. (\textbf{f-h}) Same values of magnetic field but in an longitudinal orientation ($H_x$). Note that the images do not share the same color scale (see individual color bars). For a full set of $B_z^{dc}(x,y)$ images with a transverse field orientation ($H_y$), see \href{https://youtu.be/i7FJTW6-7vQ}{supplementary movie 1}. For  a longitudinal field orientation ($H_x$), see \href{https://youtu.be/__iWGTqikUA}{supplementary movie 2}.}
    \label{fig:DC_images_field_dependant}
\end{figure}

To quantify better  the asymmetry, we plot the difference in absolute value of the critical current in each direction ($\Delta I_c=\lvert I_c^+\rvert-\lvert I_c^-\rvert$) versus applied magnetic field (Fig. \ref{fig:DC_images_field_dependant}a and methods). A further clarification on the definition of $I_c$ in this context is presented in supplementary note 1. For $H_y$, 3 states are visible. $\Delta I_c$ is positive when the magnetization is in the $+y$ direction, $\Delta I_c$ is negative when the magnetization is in the $-y$ direction, and $\Delta I_c \sim 0$ around $H_c$. The values of $H_c$ (about 5 mT) revealed by the diode effect are consistent with the volumetric magnetization measurement of the unpatterned bilayer of matching thicknesses (Fig. \ref{fig:DC_images_field_dependant}b). This confirms that the direction of the vortex diode depends on the magnetization-orientation and that when the net magnetization vanishes so does the diode effect. From the images acquired for different longitudinal fields $H_x$, we expect no diode effect at any applied field; indeed we find $\Delta I_c\sim 0$ for all values of $H_x$.

We now turn to the origin of the vortex diode effect by looking at the magnetic texture of the EuS. The out-of-plane component of the static magnetic field  $B^{dc}_z(x,y)$ was acquired as a function of the applied in-plane magnetic field (Fig. \ref{fig:DC_images_field_dependant}c-e). In these SOT images, only the stray field from the sample as a result of the magnetization of the EuS is visible. For a fully magnetized sample along the $y$ direction, large stray fields are observed at the edges. For $H_y>+H_s$, the field enters on the left edge and exits on the right (see Fig. \ref{fig:DC_images_field_dependant}c). The direction of the field lines is reverted once the magnetization is reversed (Fig. \ref{fig:DC_images_field_dependant}e). At $H_y\sim H_c$, the range in $B_z^{dc}(x,y)$ is significantly smaller, by roughly a factor 12 (note the different color scale between the images), and shows a disordered structure. The observed magnetic correlation length $\xi_f$ is on the order of the SOT diameter, $\sim200$ nm, suggesting that the domain size is even smaller. The same type of images showing disordered, low-magnitude magnetic structure are observed for all applied $H_x$ (see Fig. \ref{fig:DC_images_field_dependant}f-h). In this case, the large stray field in the $z$ direction appears only at the $x$ boundaries of the sample (outside the Hall bar). Therefore, in the region of interest the field range is small in magnitude and only a disordered magnetic structure is observed.
\begin{figure}[h]
    \centering
    \includegraphics[width=1\textwidth]{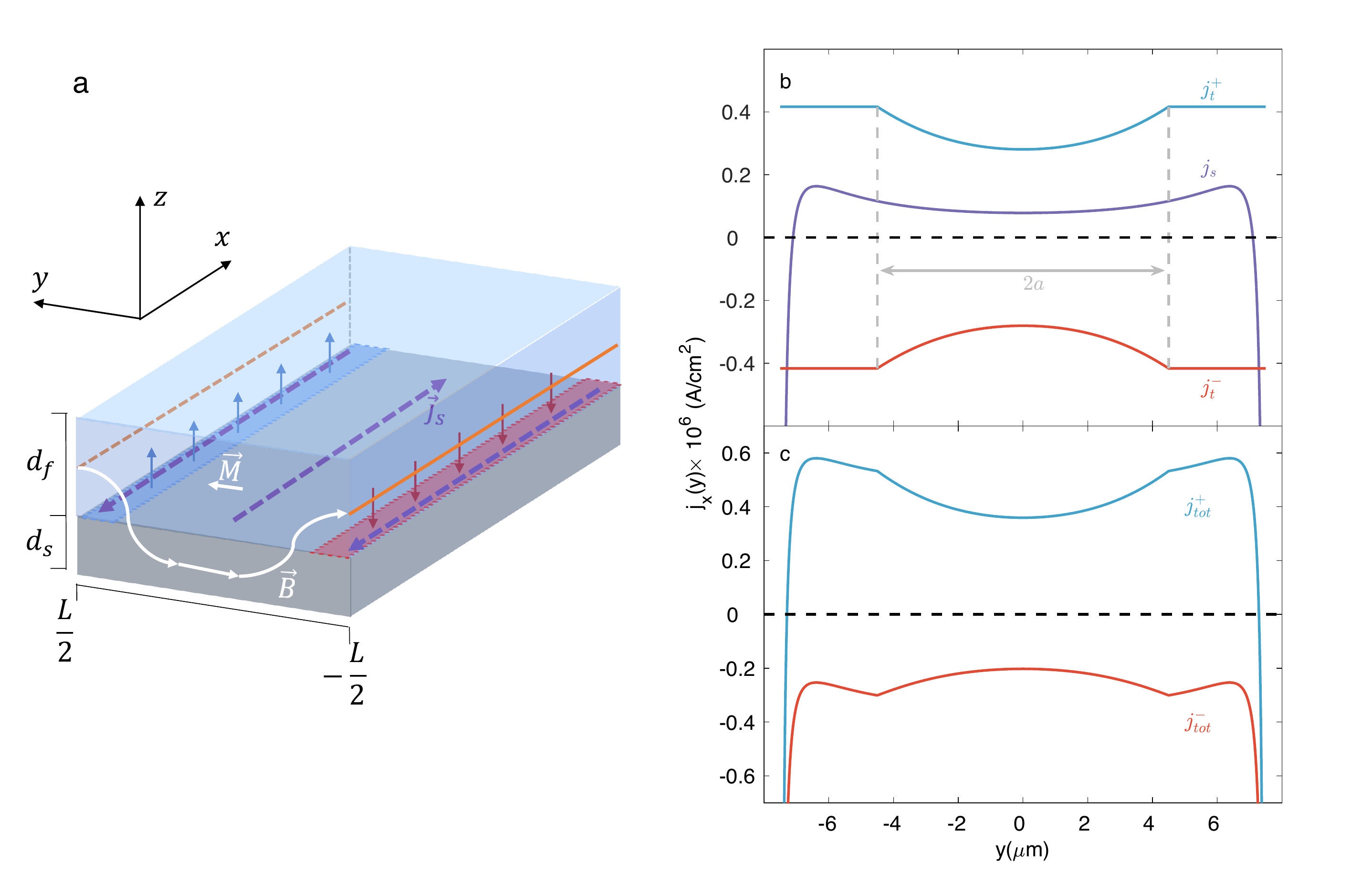}
    \caption{\small \textbf{Theoretical model} (\textbf{a}) Schematic diagram of an S(grey)/F(blue) bilayer fully magnetized along the $y$ direction. The orange lines represent auxiliary wires (at $y=\pm L/2, z=0$) with linear magnetic charge density $\pm M d_f$, which generate a stray magnetic field $\vec{H}$. This field induces a screening supercurrent $j_s(y)$ inside the S film (purple dashed lines) that generates a field in the opposite direction (red and blue arrows). The sum of the fields ($\vec{B}$) generated by the S and F layer in the $zy$ plane is represented by the white line. (\textbf{b}) Calculation of the transport current densities at opposite phases of the ac cycle (blue and red curves) along with the screening current density (purple curve). The central field-free region $2a$, is depicted by the region between the dashed grey lines.  (\textbf{c}) Total current density $j_{tot}$ (transport + screening) at opposite phases of the ac cycle. In these calculations, the value of $I_t/I_c$ (which determines the value of $a$) was set to 0.8, implying that the system is approaching the critical current.}
    \label{fig:theory_fig}
\end{figure}
\par
The quantitative observations we made in $B^{dc}_z(x,y)$ are transferred to a theoretical model in order to explain the origin of the vortex diode effect. Consider an infinite S/F bilayer strip as schematically sketched in Fig. \ref{fig:theory_fig}a. We assume that the F layer is fully magnetized along the $y$ direction and the corresponding homogeneous magnetization $4\pi M$ induces a stray field $\vec{H}$ in the $zy$ plane. The corresponding vector potential $\Vec{A}_M$ can be derived from the magnetostatic equivalent of the Poisson equation, assuming the magnetic sources are 2 infinite auxiliary wires located on the edges of the F layer (Fig. \ref{fig:theory_fig}a) \cite{buzdin_vector_potential_model}. The stray field from F is screened by the supercurrent in S. The density of the screening current, which subsequently affects vortex flow, can be found within the framework of the London approach.
\begin{equation}\label{London_relation}
    \vec{j}_s(y)=-\frac{c}{4\pi\lambda^2}\Big[ \vec{A}_\text{s}(y)+\vec{A}_\text{M}(y)+\vec{A}_0 \Big],
\end{equation}
where $\vec{A}_s$ is the vector potential induced in S and can be found from Biot-Savart's law. $\vec{A}_0$ is a gauge term that imposes $\int (\Vec{j}_s(y)\cdot\hat{x})dy=0$.
The resulting implicit equation for $\vec A_s(y)$ determines the distribution of the screening current density $j_s(y)$ in the superconducting film and can be solved iteratively \cite{buzdin_soc_vortices}.
\newline
An analytic expression for the transport current distribution can be obtained from a modification of the \textit{"Bean critical state model"} \cite{Bean_model,Zeldov_current_distribution}:
\begin{equation}\label{transport_current_density}
    j_x(y)=\begin{cases}
    \frac{2j_c}{\pi}\arctan{\left(\sqrt{\frac{(L/2)^2-a^2}{a^2-y^2}}\right)}, & \text{if $\lvert y\rvert<a$}\\
    j_c, & \text{if $a<\lvert y\rvert<L/2$} ,
    \end{cases}
\end{equation}
where we define the parameter $a=\frac{L}{2}\sqrt{1-(\frac{I_t}{I_c})^2}$, which can be interpreted as half of \textit{the central field-free region}. $I_t$ is the total transport current and $I_c$ and $J_c$ are the critical current and the critical current density, respectively.
\newline

In Fig. \ref{fig:theory_fig}b we show a comparison between the numerical calculation of the screening current density and the analytical expression for the transport current density. A more detailed derivation of the model, along with a discussion about the validity of the theoretical assumptions is provided in supplementary note 2.
\newline
It is important to note that the absolute value of the screening current density $j_s$ is larger than that of the transport current density $j_t$ along the edges but it is smaller in the bulk. Therefore, in this case, the total current density $j_{tot}^-$ (Fig. \ref{fig:theory_fig}c, red curve) would be of the same sign throughout the whole length of the device, permitting flux-flow. On the other hand, if we switch the direction of the transport current, it will tend to cancel with the screening current at the edges, resulting in a significantly lower total current in absolute value $j_{tot}^+$(blue curve). When the total current is far below the depairing limit on the edge, new vortices cannot penetrate, thus preventing flux flow. Moreover, the total current density must cross zero at two points near the edges, confining the vortices to these regions while prohibiting flux-flow. Apart from the claim that the entry barrier of the sample is overcome, another condition that needs to be fulfilled is that the barrier to leave the sample must be overcome as well. Our data suggests that the Meissner screening currents caused by the EuS has the effect of lowering both barriers. For a given saturated magnetization, vortices are repelled from the entry point and attracted by the exit point. That is what favors one direction of vortex flow. Furthermore, the existence of the diode effect, as observed both in transport and in SOT microscopy, contradicts the dominance of pinning in the bulk of the sample, as pinning should suppress the possibility of flux-flow, regardless of the entry and exit barriers. 
\section*{Discussion}

The key aspect of the model is that in the presence of stray fields from the F, the distribution of screening currents in S enhances flux flow in one direction of transport current and prevents it in the other direction. The direction of the screening currents can be reversed by reversing the magnetization and can also be significantly suppressed by setting up a magnetically disordered state in the F (i.e., by applying magnetic fields around the coercive field or by applying a field along the $x$ direction). This simple explanation captures the key aspects of our results and does not require the presence of an FFLO state. Therefore, the diode effect could be observed regardless of the SOC strength, consistent with recent data reported elsewhere\citep{moodera_diode_effect}. However, we stress that in the case where vortex dynamics is not involved, an FFLO state and large SOC strength seem to be required for the diode effect to emerge\citep{Jeon2022}. In our device, the predicted non-reciprocal critical current that triggers GHz vortex flow is observed in both local imaging techniques and global transport measurements. We therefore provide deeper insight into the underlying mechanism responsible for the superconducting diode effect. This progress should enable the development of reliable, tunable,  devices that can act, for example, as high frequency voltage rectifiers in superconducting electronics.

\section*{Methods}\label{sec:methods}
\textit{\textbf{Sample Fabrication.}}
EuS/Nb bilayers were fabricated by ulta-high vacuum electron beam evaporation at room temperature with a base pressure of better than $\sim 1\times 10^{-8}$ Torr. The main sample measured, whose results we present here, consists of a 30 nm thick layer of EuS on an 8-nm-thick layer Nb evaporated on a substrate of SiO$_2$. The EuS is capped with a 3-nm-thick layer of non-superconducting Nb. The EuS has a strong in-plane anisotropy with a saturation field of $H_s\approx10$ mT that is independent of the in-plane field angle\cite{Gomez_EuS_SMR}. The superconducting critical temperature of this sample is $\sim 5.4$ K. We present global magnetisation measurements of an unpatterned EuS/Nb device with the aforementioned thicknesses in supplementary Fig. \ref{fig:MH_MT}.
\newline
\textit{\textbf{Transport Measurements.}}
Transport measurements were carried out at 4.2 K inside a liquid helium dewar employing standard four-probe configuration, where the distances between the voltage contacts were $67.5 \mu$m. Current sweeps were performed from 0 to $\pm 0.5$ mA. A dual axis magnet consisting of standard SC coils was used to apply both OOP and IP magnetic fields. The probe itself was rotated in the sample space to control the angle of applied IP field.
\newline
\textit{\textbf{SOT Fabrication.}}
The SOT was fabricated using self aligned three step thermal deposition of Pb at cryogenic temperatures, as described in ref \cite{SOT,SOT2020}. Two sized tips were used for the purpose of the measurements. The quartz tube was pulled to create a tip diameter of  $\sim250$ nm in one case and $\sim 160$ nm in other cases. The carefully adjusted deposition thicknesses resulted in SQUID's with a critical current ranging from $60-120\mu$A at zero field. The relatively large diameter tip allows for high magnetic field sensitivity and a slight asymmetry in the Josephson junctions shifts the interference pattern of the SQUID, resulting in finite magnetic field sensitivity at low applied fields \cite{SOT}. This is crucial in order to conduct the experiment at low enough perpendicular field to avoid overcrowding the sample with vortices.
\newline
\textit{\textbf{Scanning SOT measurements.}}
The sample was zero-field cooled (ZFC) from room temperature to 4.2 K, below the critical temperature ($T_c=5.4$ K, see Fig. \ref{fig:System&Current_dependance}d). An alternating current $I^{ac}_x$ at $f\approx1.1$ kHz was imposed along the $x$ axis. Simultaneously, using the SOT, the out-of-plane (OOP) component of the magnetic field that is modulated in phase with the current, $B^{ac}_z(x,y)$ is recorded. In order to generate free vortices in the sample an OOP field $\lvert \mu_0H_z\rvert=5$ mT was applied in all the measurements presented in this work.
\newline
\textit{\textbf{Data Analysis}}
In Fig. \ref{fig:DC_images_field_dependant}a, where we present the asymmetry factor $\Delta I_c$, $I_c$ is defined as the lowest current in which we measure an onset of voltage. To determine this point, we take the derivative of $V(I)$ and impose a threshold above the noise level of 20 $m\Omega$. Supplementary Fig. \ref{fig:derivative} shows the derivative of a typical $V(I)$ curve with the corresponding threshold of 20 $m\Omega$ that was used to obtain the data in Fig. \ref{fig:DC_images_field_dependant}a. It is also possible to perform this analysis by assuming a threshold on the voltage value of the curves. For this method, we impose a threshold of 2 $\mu V$. We obtain similar results using this criterion as shown in supplementary Fig. \ref{fig:voltage}.

\backmatter
\bmhead{Data Availability}
The data that supports the findings of this study are available from the corresponding author upon reasonable request.
\bmhead{Supplementary Information}\label{sup_inf}

Supplementary Movies 1-2, notes 1-2 and figure 1.
\newline
Supplementary movie 1: \href{https://youtu.be/i7FJTW6-7vQ}{https://youtu.be/i7FJTW6-7vQ}
\newline
Supplementary movie 2: \href{https://youtu.be/__iWGTqikUA}{https://youtu.be/\_\_iWGTqikUA}



\bmhead{Acknowledgments}

We thank O. Agam, D. Orgad, E. Zeldov and A. Kamra for fruitful discussions. This work was supported by the European Research Council (ERC) Foundation Grant No. 802952, and the EPSRC through the Core-to-Core International Network “Oxide Superspin” (Grant No. EP/P026311/1). The international collaboration on this work was fostered by the EU-COST Action nanocohybri CA16218 and superqumap CA21144.

\bmhead{Author Contributions}

A.G., Y.A. O.M. and J.W.A.R. conceived the experiment. A.G. realized the SOT experiment. A.G., O.M, J.W.A.R. and Y.A. analyzed and interpreted the data. A.G., A.N, and N.F. fabricated the SOT sensor. H.M., L.G. and G.Y. fabricated the sample and performed the volumetic magnetization measurements. V.P. and A.B. conceived the theoretical model. A.G., O.M., J.W.A.R. and Y.A. wrote the manuscript with input from all the coauthors.

\bibliography{references}


\begin{thebibliography}{38}
\ifx \bisbn   \undefined \def \bisbn  #1{ISBN #1}\fi
\ifx \binits  \undefined \def \binits#1{#1}\fi
\ifx \bauthor  \undefined \def \bauthor#1{#1}\fi
\ifx \batitle  \undefined \def \batitle#1{#1}\fi
\ifx \bjtitle  \undefined \def \bjtitle#1{#1}\fi
\ifx \bvolume  \undefined \def \bvolume#1{\textbf{#1}}\fi
\ifx \byear  \undefined \def \byear#1{#1}\fi
\ifx \bissue  \undefined \def \bissue#1{#1}\fi
\ifx \bfpage  \undefined \def \bfpage#1{#1}\fi
\ifx \blpage  \undefined \def \blpage #1{#1}\fi
\ifx \burl  \undefined \def \burl#1{\textsf{#1}}\fi
\ifx \doiurl  \undefined \def \doiurl#1{\url{https://doi.org/#1}}\fi
\ifx \betal  \undefined \def \betal{\textit{et al.}}\fi
\ifx \binstitute  \undefined \def \binstitute#1{#1}\fi
\ifx \binstitutionaled  \undefined \def \binstitutionaled#1{#1}\fi
\ifx \bctitle  \undefined \def \bctitle#1{#1}\fi
\ifx \beditor  \undefined \def \beditor#1{#1}\fi
\ifx \bpublisher  \undefined \def \bpublisher#1{#1}\fi
\ifx \bbtitle  \undefined \def \bbtitle#1{#1}\fi
\ifx \bedition  \undefined \def \bedition#1{#1}\fi
\ifx \bseriesno  \undefined \def \bseriesno#1{#1}\fi
\ifx \blocation  \undefined \def \blocation#1{#1}\fi
\ifx \bsertitle  \undefined \def \bsertitle#1{#1}\fi
\ifx \bsnm \undefined \def \bsnm#1{#1}\fi
\ifx \bsuffix \undefined \def \bsuffix#1{#1}\fi
\ifx \bparticle \undefined \def \bparticle#1{#1}\fi
\ifx \barticle \undefined \def \barticle#1{#1}\fi
\bibcommenthead
\ifx \bconfdate \undefined \def \bconfdate #1{#1}\fi
\ifx \botherref \undefined \def \botherref #1{#1}\fi
\ifx \url \undefined \def \url#1{\textsf{#1}}\fi
\ifx \bchapter \undefined \def \bchapter#1{#1}\fi
\ifx \bbook \undefined \def \bbook#1{#1}\fi
\ifx \bcomment \undefined \def \bcomment#1{#1}\fi
\ifx \oauthor \undefined \def \oauthor#1{#1}\fi
\ifx \citeauthoryear \undefined \def \citeauthoryear#1{#1}\fi
\ifx \endbibitem  \undefined \def \endbibitem {}\fi
\ifx \bconflocation  \undefined \def \bconflocation#1{#1}\fi
\ifx \arxivurl  \undefined \def \arxivurl#1{\textsf{#1}}\fi
\csname PreBibitemsHook\endcsname

\bibitem{Bergeret2001}
\begin{barticle}
\bauthor{\bsnm{Bergeret}, \binits{F.S.}},
\bauthor{\bsnm{Volkov}, \binits{A.F.}},
\bauthor{\bsnm{Efetov}, \binits{K.B.}}:
\batitle{Long-range proximity effects in superconductor-ferromagnet
  structures}.
\bjtitle{Phys. Rev. Lett.}
\bvolume{86},
\bfpage{4096}--\blpage{4099}
(\byear{2001}).
\doiurl{10.1103/PhysRevLett.86.4096}
\end{barticle}
\endbibitem

\bibitem{buzdin_review}
\begin{botherref}
\oauthor{\bsnm{Buzdin}, \binits{A.I.}}:
Proximity effects in superconductor-ferromagnet heterostructures.
Reviews of Modern Physics
\textbf{77}(935)
(2005)
\end{botherref}
\endbibitem

\bibitem{Eschrig2008}
\begin{barticle}
\bauthor{\bsnm{Eschrig}, \binits{M.}},
\bauthor{\bsnm{L{\"{o}}fwander}, \binits{T.}}:
\batitle{{Triplet supercurrents in clean and disordered half-metallic
  ferromagnets}}.
\bjtitle{Nature Physics}
\bvolume{4}(\bissue{2}),
\bfpage{138}--\blpage{143}
(\byear{2008}).
\doiurl{10.1038/nphys831}
\end{barticle}
\endbibitem

\bibitem{Robinson2010}
\begin{barticle}
\bauthor{\bsnm{Robinson}, \binits{J.W.A.}},
\bauthor{\bsnm{Witt}, \binits{J.D.S.}},
\bauthor{\bsnm{Blamire}, \binits{M.G.}}:
\batitle{Controlled injection of spin-triplet supercurrents into a strong
  ferromagnet}.
\bjtitle{Science}
\bvolume{329}(\bissue{5987}),
\bfpage{59}--\blpage{61}
(\byear{2010})
{\href{https://arxiv.org/abs/https://www.science.org/doi/pdf/10.1126/science.1189246}{{https://www.science.org/doi/pdf/10.1126/science.1189246}}}.
\doiurl{10.1126/science.1189246}
\end{barticle}
\endbibitem

\bibitem{Anwar2010}
\begin{barticle}
\bauthor{\bsnm{Anwar}, \binits{M.S.}},
\bauthor{\bsnm{Czeschka}, \binits{F.}},
\bauthor{\bsnm{Hesselberth}, \binits{M.}},
\bauthor{\bsnm{Porcu}, \binits{M.}},
\bauthor{\bsnm{Aarts}, \binits{J.}}:
\batitle{Long-range supercurrents through half-metallic ferromagnetic
  ${\text{cro}}_{2}$}.
\bjtitle{Phys. Rev. B}
\bvolume{82},
\bfpage{100501}
(\byear{2010}).
\doiurl{10.1103/PhysRevB.82.100501}
\end{barticle}
\endbibitem

\bibitem{Khaire2010}
\begin{barticle}
\bauthor{\bsnm{Khaire}, \binits{T.S.}},
\bauthor{\bsnm{Khasawneh}, \binits{M.A.}},
\bauthor{\bsnm{Pratt}, \binits{W.P.}},
\bauthor{\bsnm{Birge}, \binits{N.O.}}:
\batitle{Observation of spin-triplet superconductivity in co-based josephson
  junctions}.
\bjtitle{Phys. Rev. Lett.}
\bvolume{104},
\bfpage{137002}
(\byear{2010}).
\doiurl{10.1103/PhysRevLett.104.137002}
\end{barticle}
\endbibitem

\bibitem{SC_spintronics}
\begin{botherref}
\oauthor{\bsnm{Linder}, \binits{J.}},
\oauthor{\bsnm{Robinson}, \binits{J.W.A.}}:
Superconducting spintronics.
Nature Physics
\textbf{11}
(2015)
\end{botherref}
\endbibitem

\bibitem{Yao2021}
\begin{barticle}
\bauthor{\bsnm{Yao}, \binits{Y.}},
\bauthor{\bsnm{Cai}, \binits{R.}},
\bauthor{\bsnm{Yu}, \binits{T.}},
\bauthor{\bsnm{Ma}, \binits{Y.}},
\bauthor{\bsnm{Xing}, \binits{W.}},
\bauthor{\bsnm{Ji}, \binits{Y.}},
\bauthor{\bsnm{Xie}, \binits{X.-C.}},
\bauthor{\bsnm{Yang}, \binits{S.-H.}},
\bauthor{\bsnm{Han}, \binits{W.}}:
\batitle{Giant oscillatory gilbert damping in
  superconductor/ferromagnet/superconductor junctions}.
\bjtitle{Science Advances}
\bvolume{7}(\bissue{48}),
\bfpage{3686}
(\byear{2021})
{\href{https://arxiv.org/abs/https://www.science.org/doi/pdf/10.1126/sciadv.abh3686}{{https://www.science.org/doi/pdf/10.1126/sciadv.abh3686}}}.
\doiurl{10.1126/sciadv.abh3686}
\end{barticle}
\endbibitem

\bibitem{Komori2020:ScienceAdvances}
\begin{barticle}
\bauthor{\bsnm{Komori}, \binits{S.}},
\bauthor{\bsnm{Devine-Stoneman}, \binits{J.}},
\bauthor{\bsnm{Ohnishi}, \binits{K.}},
\bauthor{\bsnm{Yang}, \binits{G.}},
\bauthor{\bsnm{Devizorova}, \binits{Z.}},
\bauthor{\bsnm{Mironov}, \binits{S.}},
\bauthor{\bsnm{Montiel}, \binits{X.}},
\bauthor{\bsnm{Olthof}, \binits{L.}},
\bauthor{\bsnm{Cohen}, \binits{L.F.}},
\bauthor{\bsnm{Kurebayashi}, \binits{H.}},
\bauthor{\bsnm{Blamire}, \binits{M.G.}},
\bauthor{\bsnm{Buzdin}, \binits{A.I.}},
\bauthor{\bsnm{Robinson}, \binits{J..W.A.}}:
\batitle{Spin-orbit coupling suppression and singlet-state blocking of
  spin-triplet cooper pairs}.
\bjtitle{Science Advances}
\bvolume{7},
\bfpage{0128}
(\byear{2020}).
\doiurl{10.1126/sciadv.abe0128}
\end{barticle}
\endbibitem

\bibitem{Boost}
\begin{botherref}
\oauthor{\bsnm{Yang}, \binits{G.}},
\oauthor{\bsnm{Ciccarelli}, \binits{C.}},
\oauthor{\bsnm{Robinson}, \binits{J.W.A.}}:
Boosting spintronics with superconductivity.
APL Materials
(2021)
\end{botherref}
\endbibitem

\bibitem{PhysRevX.10.031020}
\begin{barticle}
\bauthor{\bsnm{Jeon}, \binits{K.-R.}},
\bauthor{\bsnm{Montiel}, \binits{X.}},
\bauthor{\bsnm{Komori}, \binits{S.}},
\bauthor{\bsnm{Ciccarelli}, \binits{C.}},
\bauthor{\bsnm{Haigh}, \binits{J.}},
\bauthor{\bsnm{Kurebayashi}, \binits{H.}},
\bauthor{\bsnm{Cohen}, \binits{L.F.}},
\bauthor{\bsnm{Chan}, \binits{A.K.}},
\bauthor{\bsnm{Stenning}, \binits{K.D.}},
\bauthor{\bsnm{Lee}, \binits{C.-M.}},
\bauthor{\bsnm{Eschrig}, \binits{M.}},
\bauthor{\bsnm{Blamire}, \binits{M.G.}},
\bauthor{\bsnm{Robinson}, \binits{J.W.A.}}:
\batitle{Tunable pure spin supercurrents and the demonstration of their
  gateability in a spin-wave device}.
\bjtitle{Phys. Rev. X}
\bvolume{10},
\bfpage{031020}
(\byear{2020}).
\doiurl{10.1103/PhysRevX.10.031020}
\end{barticle}
\endbibitem

\bibitem{PhysRevB.99.024507}
\begin{barticle}
\bauthor{\bsnm{Jeon}, \binits{K.-R.}},
\bauthor{\bsnm{Ciccarelli}, \binits{C.}},
\bauthor{\bsnm{Kurebayashi}, \binits{H.}},
\bauthor{\bsnm{Cohen}, \binits{L.F.}},
\bauthor{\bsnm{Montiel}, \binits{X.}},
\bauthor{\bsnm{Eschrig}, \binits{M.}},
\bauthor{\bsnm{Komori}, \binits{S.}},
\bauthor{\bsnm{Robinson}, \binits{J.W.A.}},
\bauthor{\bsnm{Blamire}, \binits{M.G.}}:
\batitle{Exchange-field enhancement of superconducting spin pumping}.
\bjtitle{Phys. Rev. B}
\bvolume{99},
\bfpage{024507}
(\byear{2019}).
\doiurl{10.1103/PhysRevB.99.024507}
\end{barticle}
\endbibitem

\bibitem{PhysRevB.99.144503}
\begin{barticle}
\bauthor{\bsnm{Jeon}, \binits{K.-R.}},
\bauthor{\bsnm{Ciccarelli}, \binits{C.}},
\bauthor{\bsnm{Kurebayashi}, \binits{H.}},
\bauthor{\bsnm{Cohen}, \binits{L.F.}},
\bauthor{\bsnm{Komori}, \binits{S.}},
\bauthor{\bsnm{Robinson}, \binits{J.W.A.}},
\bauthor{\bsnm{Blamire}, \binits{M.G.}}:
\batitle{Abrikosov vortex nucleation and its detrimental effect on
  superconducting spin pumping in
  ${\mathrm{pt}/\mathrm{nb}/\mathrm{ni}}_{80}{\mathrm{fe}}_{20}/\mathrm{Nb}/\mathrm{Pt}$
  proximity structures}.
\bjtitle{Phys. Rev. B}
\bvolume{99},
\bfpage{144503}
(\byear{2019}).
\doiurl{10.1103/PhysRevB.99.144503}
\end{barticle}
\endbibitem

\bibitem{fulde_ferrell}
\begin{botherref}
\oauthor{\bsnm{Fulde}, \binits{P.}},
\oauthor{\bsnm{Ferrell}, \binits{R.A.}}:
Superconductivity in a strong spin-exchange field.
Physical Review
\textbf{135}(A550)
(1963)
\end{botherref}
\endbibitem

\bibitem{larkin_ovchinnikov}
\begin{botherref}
\oauthor{\bsnm{Larkin}, \binits{A.I.}},
\oauthor{\bsnm{Ovchinnikov}, \binits{Y.N.}}:
Non-uniform state of superconductors.
Zh.Eksp.Teor.Fiz.
\textbf{47}
(1964)
\end{botherref}
\endbibitem

\bibitem{Rashba_SOC}
\begin{botherref}
\oauthor{\bsnm{Rashba}, \binits{E.I.}},
\oauthor{\bsnm{Sheka}, \binits{V.I.}}:
Symmetry of energy bands in crystals of wurtzite type: Ii. symmetry of bands
  including spin-orbit interaction.
Fiz. Tverd. Tela: Collected Papers
\textbf{2}(162-76)
(1959)
\end{botherref}
\endbibitem

\bibitem{moodera_exchange_field}
\begin{barticle}
\bauthor{\bsnm{Hao}, \binits{X.}},
\bauthor{\bsnm{Moodera}, \binits{J.S.}},
\bauthor{\bsnm{Meservey}, \binits{R.}}:
\batitle{Thin-film superconductor in an exchange field}.
\bjtitle{Phys. Rev. Lett.}
\bvolume{67},
\bfpage{1342}--\blpage{1345}
(\byear{1991}).
\doiurl{10.1103/PhysRevLett.67.1342}
\end{barticle}
\endbibitem

\bibitem{moodera_symmetry_breaking}
\begin{botherref}
\oauthor{\bsnm{Hope}, \binits{M.K.}},
\oauthor{\bsnm{Amundsen}, \binits{M.}},
\oauthor{\bsnm{Suri}, \binits{D.}},
\oauthor{\bsnm{Moodera}, \binits{J.S.}},
\oauthor{\bsnm{Kamra}, \binits{A.}}:
Interfacial control of vortex-limited critical current in type-ii
  superconductor films.
Physical Review B
\textbf{104}(184512)
(2021)
\end{botherref}
\endbibitem

\bibitem{Rashba_SC_diode_theory}
\begin{barticle}
\bauthor{\bsnm{Ili}, \binits{c.S.}},
\bauthor{\bsnm{Bergeret}, \binits{F.S.}}:
\batitle{Theory of the supercurrent diode effect in rashba superconductors with
  arbitrary disorder}.
\bjtitle{Phys. Rev. Lett.}
\bvolume{128},
\bfpage{177001}
(\byear{2022}).
\doiurl{10.1103/PhysRevLett.128.177001}
\end{barticle}
\endbibitem

\bibitem{asymmetric_Ic_barber}
\begin{botherref}
\oauthor{\bsnm{Papon}, \binits{A.}},
\oauthor{\bsnm{Senapatia}, \binits{K.}},
\oauthor{\bsnm{Barber}, \binits{Z.H.}}:
Asymmetric critical current of niobium microbridges with ferromagnetic stripe.
Applied Physics Letters
\textbf{93}(172507)
(2008)
\end{botherref}
\endbibitem

\bibitem{Jeon2022}
\begin{barticle}
\bauthor{\bsnm{Jeon}, \binits{K.-R.}},
\bauthor{\bsnm{Kim}, \binits{J.-K.}},
\bauthor{\bsnm{Yoon}, \binits{J.}},
\bauthor{\bsnm{Jeon}, \binits{J.-C.}},
\bauthor{\bsnm{Han}, \binits{H.}},
\bauthor{\bsnm{Cottet}, \binits{A.}},
\bauthor{\bsnm{Kontos}, \binits{T.}},
\bauthor{\bsnm{Parkin}, \binits{S.S.P.}}:
\batitle{{Zero-field polarity-reversible Josephson supercurrent diodes enabled
  by a proximity-magnetized Pt barrier}}.
\bjtitle{Nature Materials}
\bvolume{21}(\bissue{9}),
\bfpage{1008}--\blpage{1013}
(\byear{2022}).
\doiurl{10.1038/s41563-022-01300-7}
\end{barticle}
\endbibitem

\bibitem{moodera_diode_effect}
\begin{barticle}
\bauthor{\bsnm{Hou}, \binits{Y.}},
\bauthor{\bsnm{Kamra}, \binits{A.}},
\bauthor{\bsnm{Fu}, \binits{L.}},
\bauthor{\bsnm{Lee}, \binits{P.}},
\bauthor{\bsnm{Moodera}, \binits{J.}}, \betal:
\batitle{Ubiquitous superconducting diode effect in superconductor thin films}.
\bjtitle{arXiv}
(\byear{2022}).
\doiurl{10.48550/arXiv.2205.09276}
\end{barticle}
\endbibitem

\bibitem{Suri2022}
\begin{botherref}
\oauthor{\bsnm{Suri}, \binits{D.}},
\oauthor{\bsnm{Kamra}, \binits{A.}},
\oauthor{\bsnm{Meier}, \binits{T.}},
\oauthor{\bsnm{Kronseder}, \binits{M.}},
\oauthor{\bsnm{Belzing}, \binits{W.}},
\oauthor{\bsnm{Back}, \binits{C.}},
\oauthor{\bsnm{Strunk}, \binits{C.}}:
Non-reciprocity of vortex-limited critical current in conventional
  superconducting micro-bridges.
Appl. Phys. Lett.
\textbf{121}
(2022).
\doiurl{10.48550/arXiv.2209.05754}
\end{botherref}
\endbibitem

\bibitem{josephson_diode_effect}
\begin{botherref}
\oauthor{\bsnm{Pal}, \binits{B.}},
\oauthor{\bsnm{Chakraborty}, \binits{A.}},
\oauthor{\bsnm{Sivakumar}, \binits{P.K.}}, et al.:
Josephson diode effect from cooper pair momentum in a topological semimetal.
Nature Physics
(2022)
\end{botherref}
\endbibitem

\bibitem{SC_diode_effect}
\begin{botherref}
\oauthor{\bsnm{Ando}, \binits{F.}},
\oauthor{\bsnm{Miyasaka}, \binits{Y.}},
\oauthor{\bsnm{Li}, \binits{T.}}, et al.:
Observation of superconducting diode effect.
Nature
\textbf{584}(373–376)
(2020)
\end{botherref}
\endbibitem

\bibitem{Baumgartner2022}
\begin{barticle}
\bauthor{\bsnm{Baumgartner}, \binits{C.}},
\bauthor{\bsnm{Fuchs}, \binits{L.}},
\bauthor{\bsnm{Costa}, \binits{A.}},
\bauthor{\bsnm{Reinhardt}, \binits{S.}},
\bauthor{\bsnm{Gronin}, \binits{S.}},
\bauthor{\bsnm{Gardner}, \binits{G.C.}},
\bauthor{\bsnm{Lindemann}, \binits{T.}},
\bauthor{\bsnm{Manfra}, \binits{M.J.}},
\bauthor{\bsnm{Faria~Junior}, \binits{P.E.}},
\bauthor{\bsnm{Kochan}, \binits{D.}},
\bauthor{\bsnm{Fabian}, \binits{J.}},
\bauthor{\bsnm{Paradiso}, \binits{N.}},
\bauthor{\bsnm{Strunk}, \binits{C.}}:
\batitle{Supercurrent rectification and magnetochiral effects in symmetric
  josephson junctions}.
\bjtitle{Nature Nanotechnology}
\bvolume{17}(\bissue{1}),
\bfpage{39}--\blpage{44}
(\byear{2022}).
\doiurl{10.1038/s41565-021-01009-9}
\end{barticle}
\endbibitem

\bibitem{Wu2022}
\begin{barticle}
\bauthor{\bsnm{Wu}, \binits{H.}},
\bauthor{\bsnm{Wang}, \binits{Y.}},
\bauthor{\bsnm{Xu}, \binits{Y.}},
\bauthor{\bsnm{Sivakumar}, \binits{P.K.}},
\bauthor{\bsnm{Pasco}, \binits{C.}},
\bauthor{\bsnm{Filippozzi}, \binits{U.}},
\bauthor{\bsnm{Parkin}, \binits{S.S.P.}},
\bauthor{\bsnm{Zeng}, \binits{Y.-J.}},
\bauthor{\bsnm{McQueen}, \binits{T.}},
\bauthor{\bsnm{Ali}, \binits{M.N.}}:
\batitle{The field-free josephson diode in a van der waals heterostructure}.
\bjtitle{Nature}
\bvolume{604}(\bissue{7907}),
\bfpage{653}--\blpage{656}
(\byear{2022}).
\doiurl{10.1038/s41586-022-04504-8}
\end{barticle}
\endbibitem

\bibitem{CGC_PyNb}
\begin{botherref}
\oauthor{\bsnm{Carapella}, \binits{G.}},
\oauthor{\bsnm{Granata}, \binits{V.}},
\oauthor{\bsnm{Russo}, \binits{F.}},
\oauthor{\bsnm{Costabile}, \binits{G.}}:
Bistable abrikosov vortex diode made of a py–nb ferromagnet-superconductor
  bilayer structure.
Applied Physics Letters
\textbf{94}(242504)
(2009)
\end{botherref}
\endbibitem

\bibitem{CSC_vortex_dynamics}
\begin{botherref}
\oauthor{\bsnm{Carapella}, \binits{G.}},
\oauthor{\bsnm{Sabatino}, \binits{P.}},
\oauthor{\bsnm{Costabile}, \binits{G.}}:
Asymmetry, bistability, and vortex dynamics in a finite-geometry
  ferromagnet-superconductor bilayer structure.
Physical Review B
\textbf{81}(054503)
(2010)
\end{botherref}
\endbibitem

\bibitem{Zeldov_current_distribution}
\begin{botherref}
\oauthor{\bsnm{Zeldov}, \binits{E.}},
\oauthor{\bsnm{Clem}, \binits{J.R.}},
\oauthor{\bsnm{McElfresh}, \binits{M.}},
\oauthor{\bsnm{Darwin}, \binits{M.}}:
Magnetization and transport current in thin superconducting films.
Physical Review B
\textbf{49}(14)
(1994)
\end{botherref}
\endbibitem

\bibitem{zeldov_flux_flow}
\begin{botherref}
\oauthor{\bsnm{Embon}, \binits{L.}},
\oauthor{\bsnm{Anahory}, \binits{Y.}},
\oauthor{\bsnm{Jelic}, \binits{Z.L.}},
\oauthor{\bsnm{Lachman}, \binits{E.O.}},
\oauthor{\bsnm{Myasoedov}, \binits{Y.}},
\oauthor{\bsnm{Huber}, \binits{M.E.}},
\oauthor{\bsnm{Mikitik}, \binits{G.P.}},
\oauthor{\bsnm{Silhanek}, \binits{A.V.}},
\oauthor{\bsnm{Milošević}, \binits{M.V.}},
\oauthor{\bsnm{Gurevich}, \binits{A.}},
\oauthor{\bsnm{Zeldov}, \binits{E.}}:
Imaging of super-fast dynamics and flow instabilities of superconducting
  vortices.
Nature Communications
\textbf{8}(85)
(2017)
\end{botherref}
\endbibitem

\bibitem{buzdin_vector_potential_model}
\begin{botherref}
\oauthor{\bsnm{Mironov}, \binits{S.V.}},
\oauthor{\bsnm{Buzdin}, \binits{A.I.}}:
Giant demagnetization effects induced by superconducting films.
Applied Physics Letters
\textbf{119}(102601)
(2021)
\end{botherref}
\endbibitem

\bibitem{buzdin_soc_vortices}
\begin{botherref}
\oauthor{\bsnm{Olthof}, \binits{L.A.O.}},
\oauthor{\bsnm{Montiel}, \binits{X.}},
\oauthor{\bsnm{Robinson}, \binits{J.W.}},
\oauthor{\bsnm{Buzdin}, \binits{A.I.}}:
Superconducting vortices generated via spin-orbit coupling at
  superconductor/ferromagnet interfaces.
Physical Review B
\textbf{100}(220505)
(2019)
\end{botherref}
\endbibitem

\bibitem{Bean_model}
\begin{botherref}
\oauthor{\bsnm{Bean}, \binits{C.P.}}:
Magnetization of hard superconductors.
Physical Review Letters
\textbf{8}(250)
(1962)
\end{botherref}
\endbibitem

\bibitem{Gomez_EuS_SMR}
\begin{barticle}
\bauthor{\bsnm{Gomez-Perez}, \binits{J.M.}},
\bauthor{\bsnm{Zhang}, \binits{X.-P.}},
\bauthor{\bsnm{Calavalle}, \binits{F.}},
\bauthor{\bsnm{Ilyn}, \binits{M.}},
\bauthor{\bsnm{González-Orellana}, \binits{C.}},
\bauthor{\bsnm{Gobbi}, \binits{M.}},
\bauthor{\bsnm{Rogero}, \binits{C.}},
\bauthor{\bsnm{Chuvilin}, \binits{A.}},
\bauthor{\bsnm{Golovach}, \binits{V.N.}},
\bauthor{\bsnm{Hueso}, \binits{L.E.}},
\bauthor{\bsnm{Bergeret}, \binits{F.S.}},
\bauthor{\bsnm{Casanova}, \binits{F.}}:
\batitle{Strong interfacial exchange field in a heavy metal/ferromagnetic
  insulator system determined by spin hall magnetoresistance}.
\bjtitle{Nano Letters}
\bvolume{20}(\bissue{9}),
\bfpage{6815}--\blpage{6823}
(\byear{2020}).
\doiurl{10.1021/acs.nanolett.0c02834}.
\bcomment{PMID: 32786952}
\end{barticle}
\endbibitem

\bibitem{SOT}
\begin{botherref}
\oauthor{\bsnm{Vasyukov}, \binits{D.}},
\oauthor{\bsnm{Anahory}, \binits{Y.}},
\oauthor{\bsnm{Embon}, \binits{L.}},
\oauthor{\bsnm{Halbertal}, \binits{D.}},
\oauthor{\bsnm{Cuppens}, \binits{J.}},
\oauthor{\bsnm{Neeman}, \binits{L.}},
\oauthor{\bsnm{Finkler}, \binits{A.}},
\oauthor{\bsnm{Segev}, \binits{Y.}},
\oauthor{\bsnm{Myasoedov}, \binits{Y.}},
\oauthor{\bsnm{Rappaport}, \binits{M.L.}},
\oauthor{\bsnm{Huber}, \binits{M.E.}},
\oauthor{\bsnm{Zeldov}, \binits{E.}}:
A scanning superconducting quantum interference device with single electron
  spin sensitivity.
Nature Nanotechnology
\textbf{8}(639-644)
(2013).
\doiurl{10.1038/nnano.2013.169}
\end{botherref}
\endbibitem

\bibitem{SOT2020}
\begin{barticle}
\bauthor{\bsnm{Anahory}, \binits{Y.}},
\bauthor{\bsnm{Naren}, \binits{H.R.}},
\bauthor{\bsnm{Lachman}, \binits{E.O.}},
\bauthor{\bsnm{Buhbut~Sinai}, \binits{S.}},
\bauthor{\bsnm{Uri}, \binits{A.}},
\bauthor{\bsnm{Embon}, \binits{L.}},
\bauthor{\bsnm{Yaakobi}, \binits{E.}},
\bauthor{\bsnm{Myasoedov}, \binits{Y.}},
\bauthor{\bsnm{Huber}, \binits{M.E.}},
\bauthor{\bsnm{Klajn}, \binits{R.}},
\bauthor{\bsnm{Zeldov}, \binits{E.}}:
\batitle{Squid-on-tip with single-electron spin sensitivity for high-field and
  ultra-low temperature nanomagnetic imaging}.
\bjtitle{Nanoscale}
\bvolume{12},
\bfpage{3174}--\blpage{3182}
(\byear{2020}).
\doiurl{10.1039/C9NR08578E}
\end{barticle}
\endbibitem

\bibitem{Aarts_depairing}
\begin{barticle}
\bauthor{\bsnm{Rusanov}, \binits{A.Y.}},
\bauthor{\bsnm{Hesselberth}, \binits{M.B.S.}},
\bauthor{\bsnm{Aarts}, \binits{J.}}:
\batitle{Depairing currents in superconducting films of $\mathrm{Nb}$ and
  amorphous $\mathrm{MoGe}$}.
\bjtitle{Phys. Rev. B}
\bvolume{70},
\bfpage{024510}
(\byear{2004}).
\doiurl{10.1103/PhysRevB.70.024510}
\end{barticle}
\endbibitem

\end{thebibliography}
\newpage
\renewcommand{\thefigure}{S\arabic{figure}}
\setcounter{figure}{0}
\section*{\small{Supplementary information for "Direct Observation of a Superconducting Vortex Diode"}}
\subsection*{Supplementary Figures}
\begin{figure}[h]
    \centering
    \includegraphics[width=0.9\textwidth]{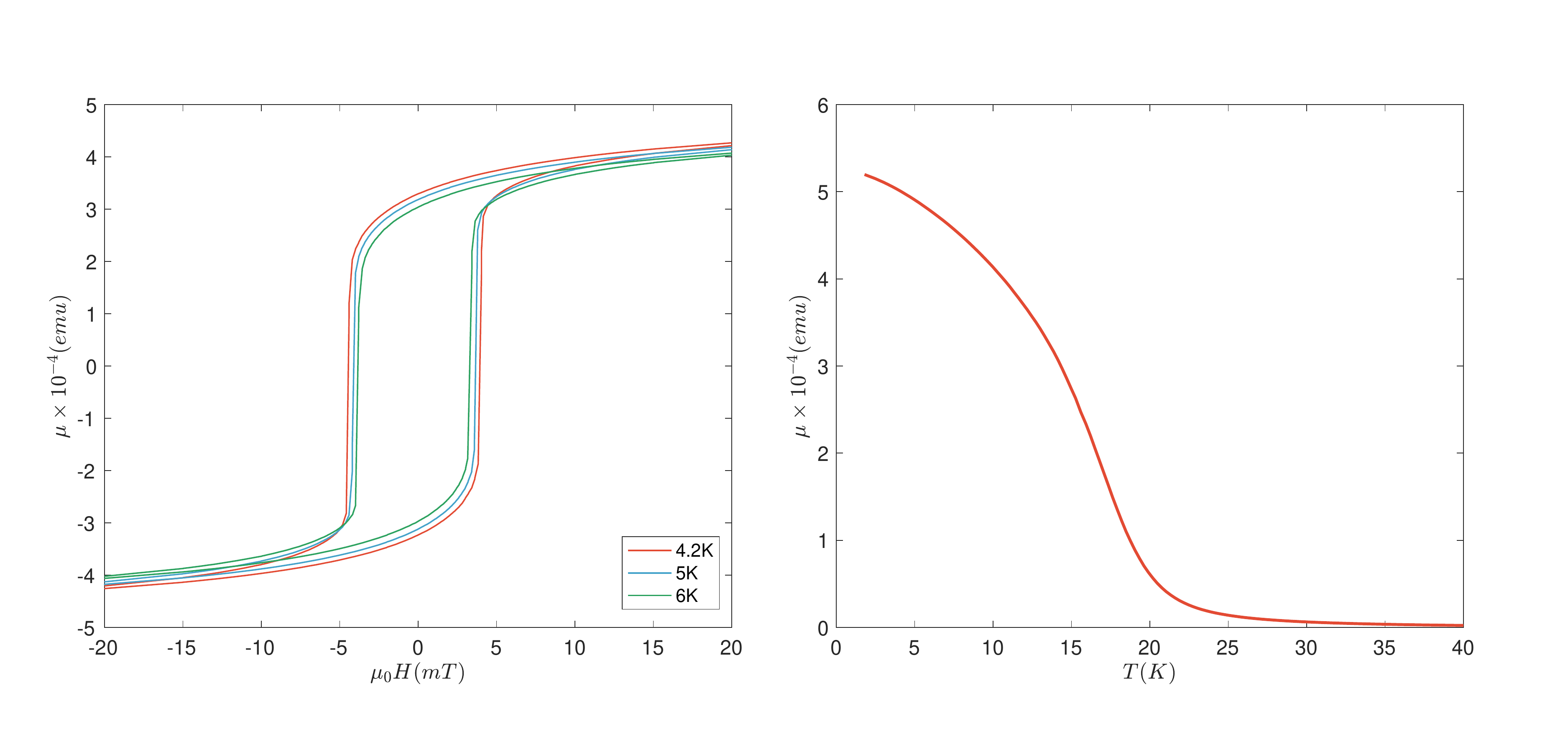}
    \caption{\textbf{Global magnetization measurements}. \textbf{a} In-plane $M(H)$ curves of an
    unpatterned EuS/Nb film with the same thicknesses as our device for different temperatures. The red curve was taken at $4.2$K, the blue curve at $5$K and the green curve at $6$K.\textbf{b} $M(T)$ curve of the same device. An in-plane field of $50$mT was applied for this measurement} 
    \label{fig:MH_MT}
\end{figure}
\subsection*{Supplementary Note I}
\begin{figure}[h]
    \centering
    \includegraphics[width=0.5\textwidth]{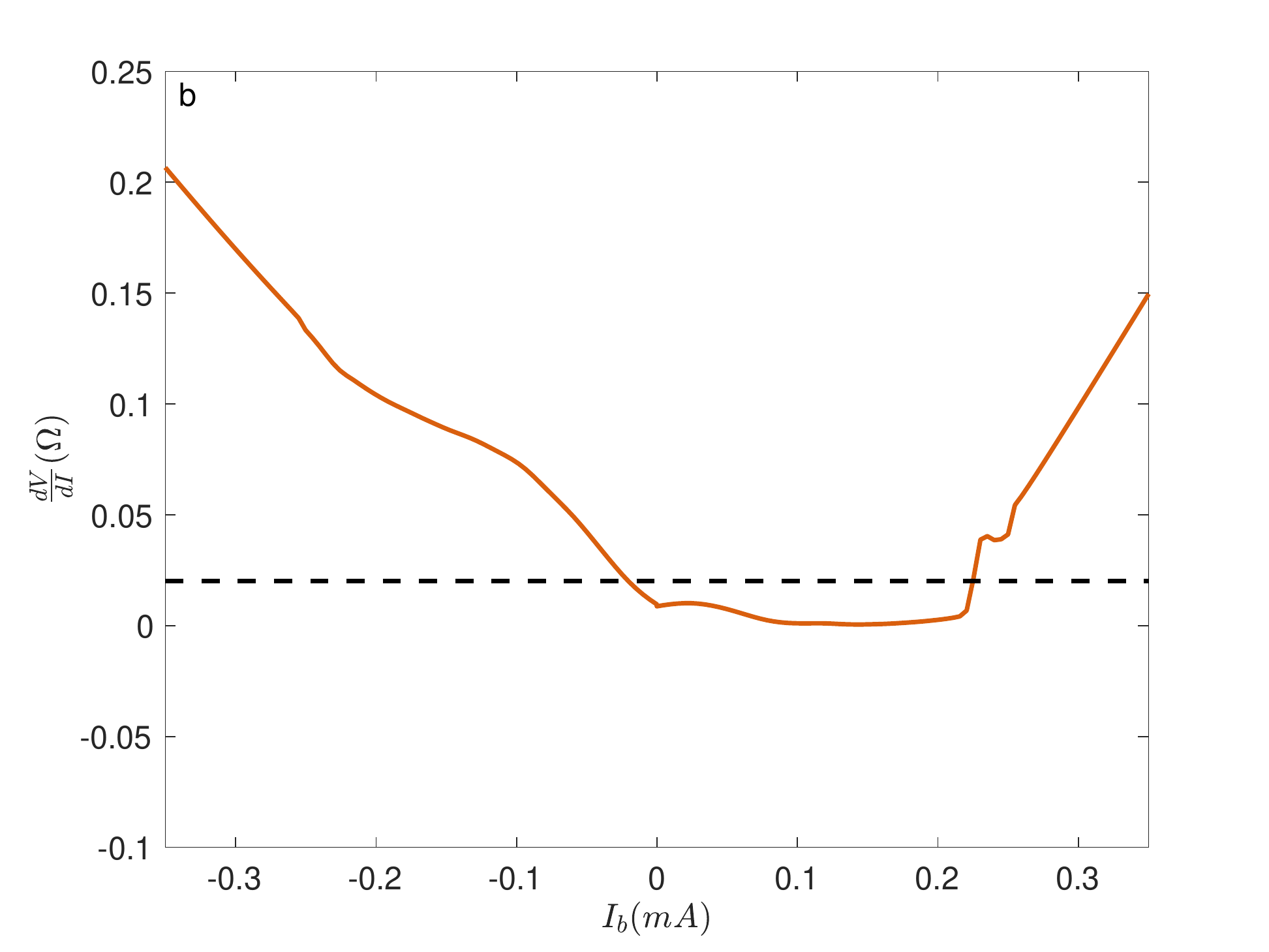}
    \caption{\textbf{Analysis of the asymmetry factor by the derivative criterion}. $\frac{dV}{dI}$ of one $I-V$ curve. A linear regression filter was applied to the raw data in order to produce a smooth derivative. The dashed line represents the threshold of $20m\Omega$}
    \label{fig:derivative}
\end{figure}

\begin{figure}[h]
    \centering
    \includegraphics[width=0.9\textwidth]{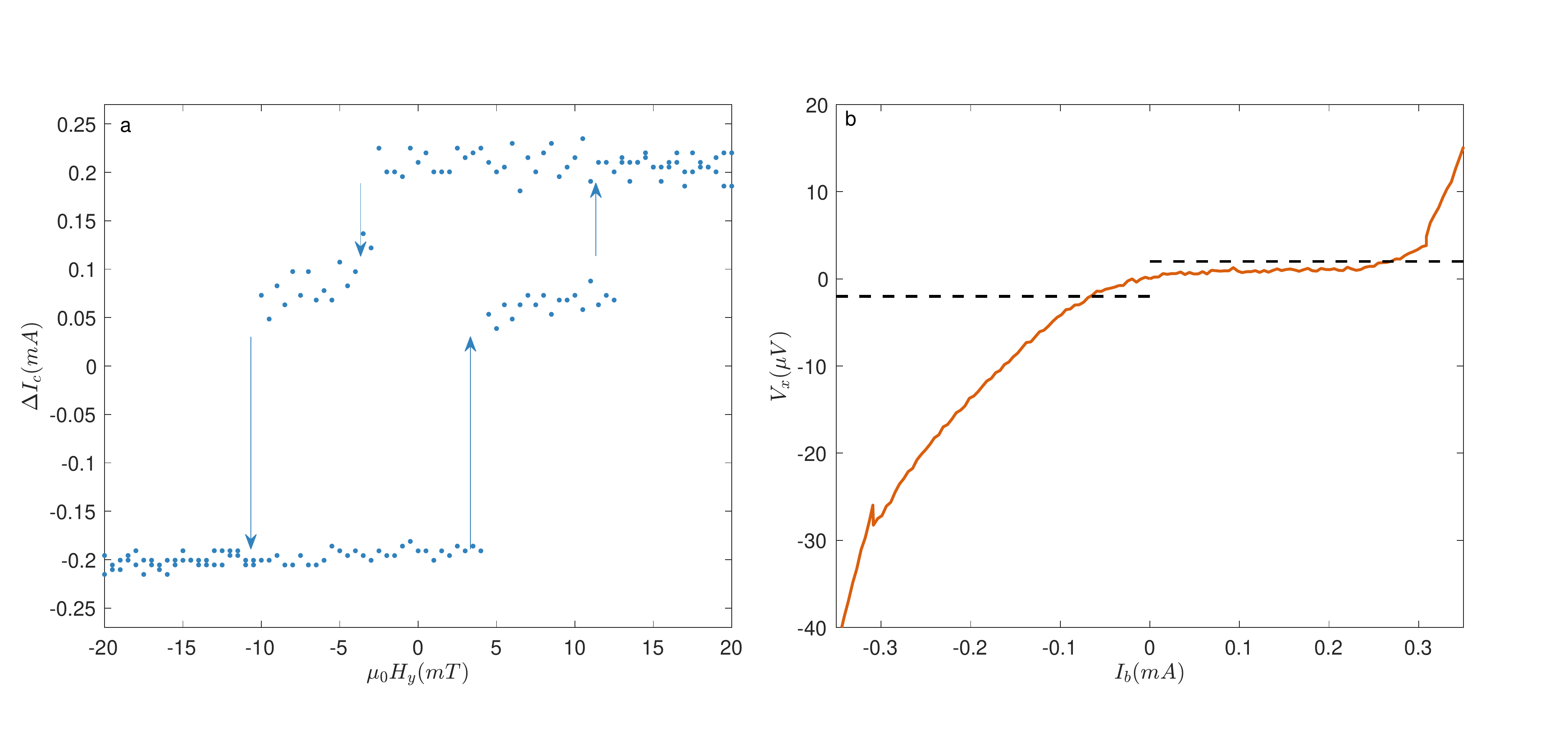}
    \caption{\textbf{Analysis of the asymmetry factor by the voltage criterion}. (\textbf{a}) Asymmetry factor $\Delta I_c=\lvert I_c^+\rvert-\lvert I_c^-\rvert$ as a function of transverse magnetic field $\mu_0 H_y$. The arrows mark the magnetic field sweep directions. (\textbf{b}) Example of one $I-V$ curve with the chosen threshold of the voltage (dashed black line).}
    \label{fig:voltage}
\end{figure}

\textbf{Defining the critical current by the I-V slope}
\newline
The following note provides a detailed explanation about the method used for defining the critical currents from which the the asymmetry factor presented in fig. 3a of the main paper was calculated. The criterion used was to impose a very small threshold on the derivative of the $I-V$ curves in order to 'detect' already a very slight increase in voltage. We note, that this criterion may not provide a well defined value of the physical critical current which gives rise to macroscopic flux flow, but rather gives a parameter that qualitatively represents the asymmetry of the transport curves. Furthermore, in order to produce clear results that focus mainly on the 3 different states ($H>H_{s+}$,$H<-H_{s-}$ and $H\approx H_c$) we smooth the raw data by applying a linear regression filter and define the threshold to be $20m\Omega$. Figure \ref{fig:derivative} shows explicitly the derivative of a single $I-V$ curve after applying the filter. $I_{c+}$ and $I_{c-}$ are obtained by the intersection of the plotted curve with the threshold of $20m\Omega$. 
\newline
\newpage
\textbf{Defining the critical current by voltage threshold}
\newline
It is also possible to perform this analysis by assuming a threshold on the voltage value in the $I-V$ curves. In this case, no smoothing of the data is necessary and we can assume a threshold of $2\mu V$. The result of this method of analysis is shown in figure \ref{fig:voltage}a. The resulting data points show exactly the same trend as those achieved by the derivative criterion (figure 3a of the main article), but they are far more scattered and can deflect the attention from the main observation.   
\newpage
\subsection*{Supplementary Note II}
\textbf{Calculation of the Screening Currents}
\newline
Here we provide details of the calculation of the screening current in the S layer induced by a stray magnetic field of the F layer where we neglect the superconducting proximity effect and electromagnetic interaction between the layers. For the limit $L \gg d_f $ the stray field in the region of the superconductor layer can be expressed with the help of a vector potential $A_{\text{M}}(y)$.
The field $\vec{H}$ can be modeled by the field induced by two infinite wires with opposite magnetic charge densities $\pm M d_f$ positioned at $y=\pm L/2$ and $z=0$ . Note here that the plane $z=0$ is chosen to pass through the middle of the F layer. We also neglect the finite thickness of the S strip and consider it as a delta layer placed in the $z=-d_f/2$ plane. The corresponding vector potential inside the superconducting film is as follows:

\begin{equation}\label{magnetic_vector_potential}
    \vec{A}_{\text{M}}(y)=-2Md_f\Bigg[ \arctan{\Bigg(\frac{2y+L}{d_f}\Bigg)}-\arctan{\Bigg(\frac{2y-L}{d_f}\Bigg)} \Bigg]\Hat{x}
\end{equation}

For a qualitative analysis consider first the case  $L \ll \lambda_\text{eff} $.
This allows one to use London's equation neglecting the contribution to the vector potential from the screening current. Thus, the supercurrent is defined through the local vector potential generated by the ferromagnet:
\begin{equation}
    \vec j_s(y)=j_s(y)\vec{x}_0=-\frac{c}{4\pi\lambda^2} \Big[ \vec A_{\text{M}}(y)+\vec A_0 \Big]
\end{equation}
where $\vec A_0$ is a gauge term which can be found from the condition  $\int_{-L/2}^{L/2}j_s(y)dy=0$. 
Thus, this gives us simple analytical expression for spatial distribution of the screening current density:
\begin{equation}
    \begin{split}
    j_s(y)/j_0=M_0\Big[\arctan{\big((2y+L)/d_f\big)}-\arctan{\big((2y-L)/d_f\big)} - \\
    2\arctan{\big(2L/d_f\big)}
    +(d_f/L)\ln{\big(1+4L^2/d^2_f\big)} \Big]
    \end{split}
\end{equation}

Here we introduced a dimensionless magnetization $M_0=4\pi M d_f/\Phi_0$, where the value $4\pi M\approx 2$T corresponds to $M_0=1$. Current density is expressed in terms of $j_0=\Phi_0c/8\pi^2d_f\lambda^2$, where in SI units for $d_f=30$nm and $\lambda=\lambda_\text{film}(0)\sim 220$nm we have $j_0\approx 17\times 10^6$ A/cm$^2$.

\

\begin{figure}[h] 
\centering
\includegraphics[width=0.5\textwidth]{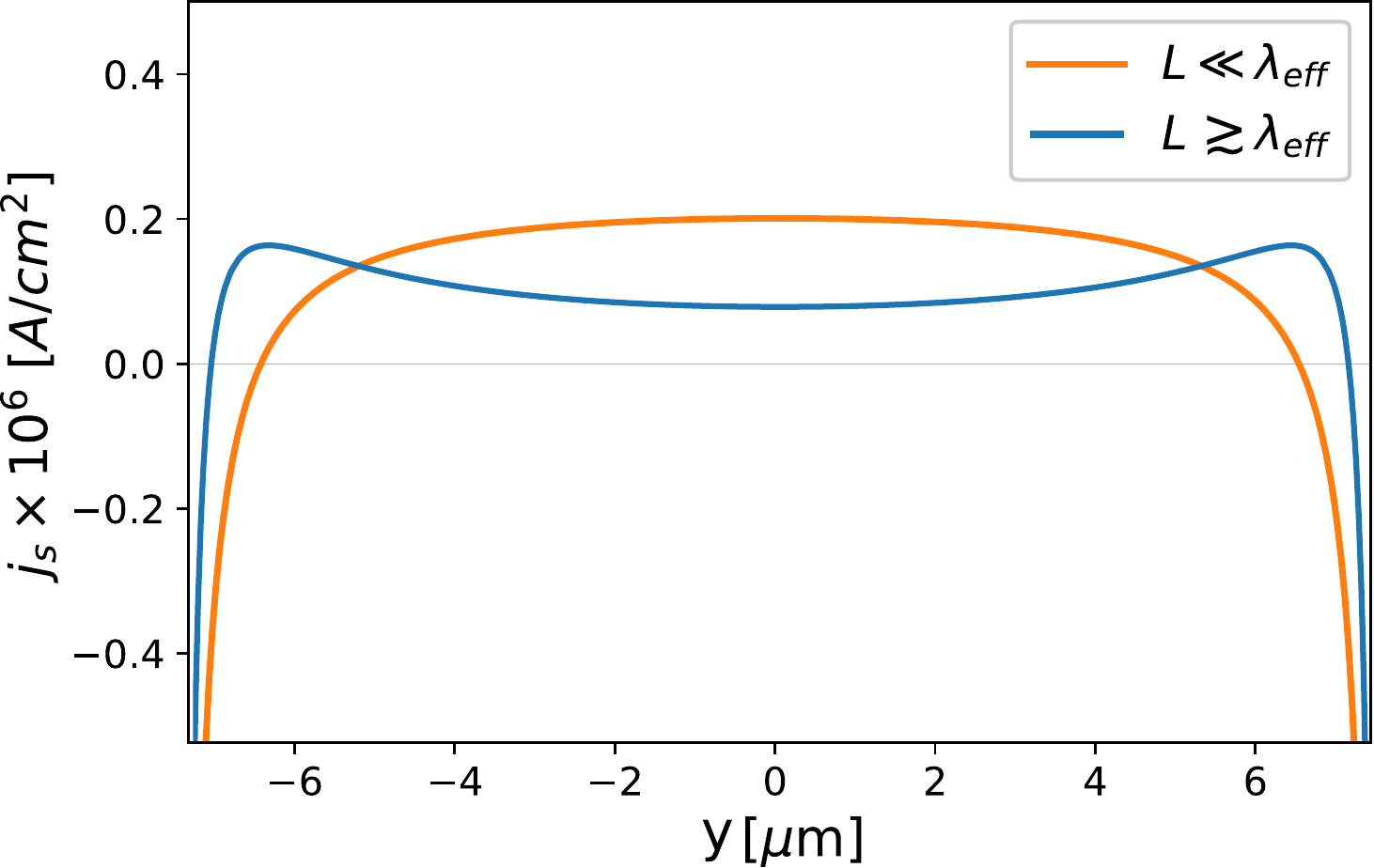} 
\caption{ {\small  Screening current distribution across the S/F bilayer strip for parameters $L/d_f=500$; $L/\lambda_\text{eff}=15$; $M_0=1$.  }}
\label{supp_js}
\end{figure}

The opposite limit $L\gtrsim\lambda_{\text{eff}}$ is of more interest in the system under consideration.
The supercurrent screening the magnetic field in the S layer satisfies the London relation
\begin{equation}
\vec j_s(y)=-\frac{c}{4\pi\lambda^2}\Big[ \vec A_\text{s}(y)+\vec A_\text{M}(y)+\vec A_0 \Big],
\end{equation}
$\vec A_0 =-\int_{-L/2}^{L/2}\big( \vec A_s+\vec A_\text{M} \big)dy/L.$ is a gauge term that imposes $\int j_x(y)dy=0$.
The vector potential induced in the superconductor can be found from Biot-Savart's law:
\begin{equation}
\vec A_s(\vec r)=\frac1c\int\frac{d \vec j_s}{R}=\frac1c\int\frac{\vec j_s(\vec{r'})d^3r'}{\lvert \vec{r}-\vec{r'}\rvert}
\end{equation}
In the case of the thin superconducting film one has:
\begin{equation}
\vec A_s(x,y)
=\frac{d_s}{c}\int_{-L/2}^{L/2}dy'\int_{-l}^{l}dx'\frac{ \vec j_s(x',y')}{\sqrt{(x-x')^2+(y-y')^2)}}
\end{equation}
Consider a strip that is infinite in $x$ direction with the length $l\rightarrow \infty$ in the presence of the supercurrent $\vec j_s=j_s(y)\vec x_0$. 
After integration the vector potential gains a simple form:
\begin{equation}
A_s(y)=-\frac{2d_s}{c}\int_{-L/2}^{-L/2}\ln{\lvert y-y'\rvert}j_s(y')dy'
\end{equation}
The last expression can be rewritten as an implicit equation for $A_s$:
\begin{equation}
\begin{split}
A_s(y)=\frac{d_s}{2\pi\lambda^2}\int_{-\frac L2}^{-\frac L2}\ln{\lvert y-y'\rvert}\Big( A_s(y')+A_\text{M}(y') \Big)dy'-\\
\frac{d_s}{2\pi\lambda^2}\frac{1}{L}\int_{-\frac L2}^{-\frac L2}\Big( A_s(y'')+A_\text{M}(y'') \Big)dy''\int_{-\frac L2}^{-\frac L2}\ln{\lvert y-y'\rvert}dy'
\end{split}
\end{equation}
This equation can be solved iteratively, starting with the ansatz $A_s(y)=0$.
\newline
The current distribution for both cases is shown in Fig. \ref{supp_js}. It is clearly seen that a simple analytical approach for $L\ll \lambda_\text{eff}$ qualitatively gives reasonable profile of the supercurrent. 

\par
\newpage
\textbf{Transport Current Distribution}
\newline
Here we present the analytic expression for the transport current distribution as obtained from a modification of the \textit{"Bean critical state model"} \cite{Bean_model,Zeldov_current_distribution}: 
\begin{equation}\label{transport_current_density}
    j_x(y)=\begin{cases}
    \frac{2j_c}{\pi}\arctan{\left(\sqrt{\frac{(L/2)^2-a^2}{a^2-y^2}}\right)}, & \text{if $\lvert y\rvert<a$}\\
    j_c, & \text{if $a<\lvert y\rvert<L/2$} ,
    \end{cases}
\end{equation}
where we define the parameter $a=\frac{L}{2}\sqrt{1-(\frac{I_t}{I_c})^2}$, which can be interpreted as half of \textit{the central field-free region}. $I_t$ is the total transport current and $I_c$ and $J_c$ are the critical current and the critical current density, respectively.
In the calculation of the transport current distribution we have chosen the magnitude of the transport current to be close to what was used in the experiment ($I_t=0.4$ mA and $I_c=0.5$ mA), but it is evident that the model shows the same qualitative behavior for a wide range of current values.
\newline

\par
We would like to elaborate now the sharp increase in the screening current density in the vicinity of the sample edge (supplemetary fig. \ref{supp_js}). The reason for these peaks is due to the large value of the ratio $L/d_f=500$. It is important to note that the model is not valid in the range $\lvert y\pm L/2\rvert<d_f$, which requires the introduction of a cutoff length scale for further investigation of $j_s$ in this range. However, since the current value $j_s$ at a distance $\sim d_f$ from the edge does not exceed the depairing current, $j_d\sim 200\times 10^6 A/cm^2$ \cite{Aarts_depairing}, this cutoff can be omitted for the purpose of our model. 
\newline
It is also necessary to discuss the influence of vortices on the screening current density. The fact that the current $j_s$ is large in the vicinity of the edges means that the stray field of the F layer may be able to induce vortices in this region. The zero-crossing of the current creates a potential well for a vortex, which results in the appearance of an equilibrium chain of vortices along the strip edge. Due to this confining potential, the large currents on the edge do not contradict the critical current $j_c$ defined within the dynamic nature of the Bean model\cite{Bean_model,Zeldov_current_distribution}. Taking into account also the vortex supercurrent, $j_v$, would further reduce the sharp peaks along the edges, but due to the localized nature of the vortex chain, in the stationary regime the bulk value of the current (in the middle of the strip) should remain mostly unperturbed.
For the aforementioned reasons, introducing a cutoff and/or adding $j_v$ to the total current density would smooth the sharp peaks along the edge, but would not change the qualitative predictions of the model. In particular, the result that we get a different sign of $j_{tot}$ in the bulk of the strip for opposite directions of $j_t$ (which explains the diode effect) is robust to these additional considerations in the model.

\end{document}